\RequirePackage{ifpdf}
\ifpdf 
\documentclass[pdftex]{sigma}
\else
\documentclass{sigma}
\fi

\begin{document}

\newcommand{\para}{\parallel}
\newcommand{\pad}{\partial}
\newcommand{\nn}{\nonumber}
\newcommand{\la}{\leftarrow}
\newcommand{\ra}{\rightarrow}
\newcommand{\lgla}{\longleftarrow}
\newcommand{\lgra}{\longrightarrow}
\newcommand{\La}{\Leftarrow}\newcommand{\Ra}{\Rightarrow}
\newcommand{\Lra}{\Leftrightarrow}
\newcommand{\Lgla}{\Longleftarrow}
\newcommand{\Lgra}{\Longrightarrow}
\newcommand{\bm}{\boldmath}
\newcommand{\lan}{\langle}
\newcommand{\ran}{\rangle}
\renewcommand{\a}{\alpha}
\renewcommand{\b}{\beta}
\newcommand{\g}{\gamma}
\newcommand{\G}{\Gamma}
\renewcommand{\d}{\delta}
\newcommand{\eps}{\epsilon}
\newcommand{\Th}{\Theta}
\newcommand{\s}{\sigma}
\newcommand{\lam}{\lambda}
\newcommand{\D}{\Delta}
\newcommand{\vare}{\varepsilon}
\newcommand{\pr}{\prime}
\newcommand{\ro}{\rho}
\newcommand{\nab}{\nabla}
\newcommand{\m}{\mu}
\newcommand{\n}{\nu}
\newcommand{\Sg}{\Sigma}
\newcommand{\p}{\pi}
\newcommand{\R}{I\!\!R}
\newcommand{\om}{\omega}
\newcommand{\Om}{\Omega}
\newcommand{\ze}{\zeta}
\newcommand{\vart}{\vartheta}
\newcommand{\tri}{\triangle}
\newcommand{\f}{\frac}
\newcommand{\iny}{\infty}
\newcommand{\pro}{\propto}

\renewcommand{\thefootnote}{$\star$}

\renewcommand{\PaperNumber}{083}

\FirstPageHeading

\ShortArticleName{Non-Hermitian Quantum Mechanics with
Minimal Length Uncertainty}

\ArticleName{Non-Hermitian Quantum Mechanics\\ with
Minimal Length Uncertainty\footnote{This paper is a
contribution to the Proceedings of the 5-th Microconference
``Analytic and Algebraic Me\-thods~V''. The full collection is
available at
\href{http://www.emis.de/journals/SIGMA/Prague2009.html}{http://www.emis.de/journals/SIGMA/Prague2009.html}}}

\Author{T.K. JANA~$^\dag$ and P. ROY~$^\ddag$}

\AuthorNameForHeading{T.K. Jana and P. Roy}

\Address{$^\dag$~Department of Mathematics, R.S. Mahavidyalaya, Ghatal 721212, India}
\EmailD{\href{mailto:tapasisi@gmail.com}{tapasisi@gmail.com}}

\Address{$^\ddag$~Physics and Applied Mathematics Unit, Indian Statistical Institute, Kolkata-700108, India}
\EmailD{\href{mailto:pinaki@isical.ac.in}{pinaki@isical.ac.in}}

\ArticleDates{Received June 30, 2009, in f\/inal form August 10, 2009;  Published online August 12, 2009}

\Abstract{We study non-Hermitian quantum mechanics in the presence of a minimal length. In particular we obtain exact solutions of a non-Hermitian displaced harmonic oscillator and the Swanson model with minimal length uncertainty. The spectrum in both the cases are found to be real. It is also shown that the models are $\eta$ pseudo-Hermitian and the metric operator is found explicitly in both the cases.}


\Classification{81Q05; 81S05}

\renewcommand{\thefootnote}{\arabic{footnote}}
\setcounter{footnote}{0}

\section{Introduction}

In recent years there have been growing interest on quantum systems with a minimal length \cite{kempf1,kempf2,kempf3,kempf4}. There are quite a few reasons for this. For example, the concept of minimal length has found applications in quantum gravity \cite{garay}, perturbative string theory \cite{gross2}, black holes \cite{magg} etc. Exact as well as perturbative solutions of various non relativistic quantum mechanical systems, e.g., harmonic oscillator \cite{kempf2,kempf3,chang,dadic,gemba}, Coulomb problem \cite{brau,tk,yao,chang1}, Pauli equation \cite{nou} etc., have been obtained in the presence of minimal length. Exact solutions of relativistic models like the Dirac oscillator have also been obtained \cite{tk1,nou1}. A novel approach based on momentum space supersymmetry was also used to obtain exact solutions of a number of problems \cite{tk1,tk2,spector}.

On the other hand, since the work of Bender et al.~\cite{bender} non-Hermitian quantum systems have been studied extensively over the past few years\footnote{See \url{http://gemma.ujf.cas.cz/~znojil/conf/}.}.
Many of these models, especially the $\cal{PT}$ symmetric and the $\eta$ pseudo-hermitian ones admit real spectrum in spite of being non-Hermitian. Recently some possible applications of non-Hermitian quantum mechanics have also been suggested \cite{markis,moi}. However all these studies have been made in the context of point particles. Here our aim is to examine non-Hermitian quantum mechanics in the presence of a~minimal length. In particular we shall obtain exact solutions of a displaced harmonic oscillator with a complex coupling and the Swanson model~\cite{swanson}. It will be shown that in both the cases the spectrum is entirely real (subject to the parameters in the later case satisfying some constraints depending on the minimal length) and both the models are in fact $\eta$ pseudo-Hermitian. Explicit representation of the  metric will also be obtained in both the cases. The organization of the paper is as follows. In Section~\ref{QM} we present a few results concerning quantum mechanics with minimal length uncertainty. In~Section~\ref{dho} we present exact solutions of the displaced harmonic oscillator problem. Section \ref{swanson} contains exact solutions of the Swanson model. In Section~\ref{pseudo} we discuss $\eta$ pseudo-Hermiticity of the models and finally Section~\ref{con} is devoted to a discussion.

\section{Quantum mechanics with minimal length uncertainty}\label{QM}

In one dimensional quantum mechanics with a minimal length the canonical commutation relation between $\hat{x}$ and $\hat{p}$ is modified and reads \cite{kempf2}
\begin{gather}
[{\hat x},{\hat p}] = i\hbar\big(1+\beta p^2\big),\label{cano}
\end{gather}
where $\b$ is a small parameter. A representation of $\hat{x}$ and $\hat{p}$ which realizes (\ref{cano}) is given by \cite{kempf2}
\begin{gather}
{\hat x} = i\hbar\left[\big(1+\beta p^2\big)\f{\partial}{\partial p}+\gamma p\right],\qquad {\hat p} = p. \label{rep1}
\end{gather}
From (\ref{cano}) and (\ref{rep1}) it can be shown that
\begin{gather}
\Delta{\hat x}\Delta {\hat p} \geq \f{\hbar}{2}\big[1+\beta (\Delta {\hat p})^2\big],\label{uncer}
\end{gather}
where in obtaining (\ref{uncer}) we have taken $\langle p\rangle =0$. Thus the standard Heisenberg uncertainty relation (corresponding to $\b\ra 0$) is modified and it follows that there is UV/IR mixing. Furthermore from (\ref{uncer}) it follows that there also exist a minimal length given by
\begin{gather*}
(\Delta {\hat x})_{\min} = \hbar \sqrt{\beta}.
\end{gather*}
In the space where position $({\hat x})$ and momentum $({\hat p})$ are given by (\ref{rep1}) the associated scalar product is defined by
\begin{gather}
\langle \phi(p)|\psi(p)\rangle = \int \f{\phi^*(p)\psi(p)}{(1+\beta p^2)^{1-\f{\gamma}{\beta}}}\, dp.\label{scalar1}
\end{gather}

\section{Non-Hermitian displaced harmonic oscillator}\label{dho}

The Schr\"odinger equation for the displaced oscillator is given by
\begin{gather}
H\psi(p) = E\psi(p),\qquad H = \f{1}{2\mu}{\hat p}^2 + \f{1}{2}\mu \omega^2{\hat x}^2 + i\lam{\hat x},\label{sch1}
\end{gather}
where $\lam$ is a real constant. Now using (\ref{scalar1}) it can be shown that
\begin{gather*}
H \neq H^\dagger,
\end{gather*}
so that $H$ is non-Hermitian.
Then we use (\ref{rep1}) to write the Schr\"odinger equation (\ref{sch1}) in momentum space as
\begin{gather}
\left[-f(p) \f{d^2}{dp^2} + g(p)\f{d}{dp}+h(p)\right]\psi(p) = \eps \psi(p),\label{sch2}
\end{gather}
where $f(p)$, $g(p)$, $h(p)$ and $\eps$ are given by
\begin{gather}
f(p)  =  \big(1+\beta p^2\big)^2,\qquad
g(p)  =  -2\big(1+\beta p^2\big)\left[(\gamma+\beta)p+\f{\lam}{\mu\hbar \omega^2}\right],\nonumber\\
h(p)  =   \left[\f{1}{{\hbar}^2\mu^2\omega^2}-\gamma(\beta+\gamma)\right]p^2-\f{2\lam\g}{\hbar\mu\om^2}p,\qquad
\eps  =   \f{2E}{{\hbar}^2\mu\omega^2}+\g.\label{fgh}
\end{gather}
It is now necessary to solve equation~(\ref{sch2}). To this end we perform a simultaneous change of wave function as well as the independent variable:
\begin{gather}
\psi(p) = \rho(p)\phi(p), \qquad q = \int \f{1}{\sqrt{f(p)}}\, dp, \label{t}
\end{gather}
where
\begin{gather}
\rho(p) = e^{\int \chi(p)\,dp},\qquad \chi(p) =  \f{f^\prime+2g}{4f}. \label{rho}
\end{gather}
Using the transformation (\ref{t}) we obtain from (\ref{sch2})
\begin{gather}
\left[-\f{d^2}{dq^2} + V(q)\right]\phi(q) = \eps\phi(q), \label{sch3}
\end{gather}
where $V(q)$ is given by
\begin{gather*}
V(q) = \left[\f{4g^2+3{f^\prime}^2+8gf^\prime}{16f}-\f{f^{\prime\prime}}{4}-\f{g^\prime}{2} + h(p)\right]_q.
\end{gather*}
It is easy to see that (\ref{sch3}) is a standard Schr\"odinger equation in the variable $q$ and $V(q)$ is the corresponding potential. In the present case we obtain on using (\ref{fgh})
\begin{gather}
q = \f{1}{\sqrt{\beta}}\, {\rm tan}^{-1}\big(\sqrt{\beta}p\big),\qquad -\f{\pi}{2\sqrt{\beta}}<q<\f{\pi}{2\sqrt{\beta}},
\nonumber\\
V(q) = \f{{\rm sec}^2(\sqrt{\beta}q)}{{\hbar}^2\mu^2\omega^2\beta} +\f{\lam^2}{{\hbar}^2\mu^2\omega^4}-\f{1}{{\hbar}^2\mu^2\omega^2\b}+\gamma. \label{pot1}
\end{gather}
The potential $V(q)$ given above is a standard solvable potential. The energy eigenvalues and the wave functions are given by \cite{khare}
\begin{gather*}
\eps_n  =  \big(A+n\sqrt{\b}\,\big)^2+\f{\lam^2}{{\hbar}^2\mu^2\omega^4}-\f{1}{{\hbar}^2\mu^2\omega^2\b}+\gamma,\qquad n=0,1,2,\dots,\\
\phi_n(q)  = N_n \big[\cos \big(q\sqrt{\b}\,\big)\big]^{\f{A}{\sqrt{\b}}} P_n^{\big(\f{A}{\sqrt{\b}}-\f{1}{2},\f{A}{\sqrt{\b}}-\f{1}{2}\big)}\big(\sin \big(q\sqrt{\b}\,\big)\big),\\
A  =   \f{\sqrt{\b}+\sqrt{\b+\f{4}{\hbar^2\mu^2\omega^2\b}}}{2},
\end{gather*}
where $N_n$ are normalization constants and $P_n^{(r,s)}(z)$ denotes Jacobi polynomials.

So from (\ref{fgh}) and (\ref{t}) we finally obtain ($n=0,1,2,\dots$)
\begin{gather}
E_n = {\hbar}\omega\left[\f{\beta\hbar\omega\mu}{2}\left(n^2+n+\f{1}{2}\right)
+\left(n+\f{1}{2}\right)\sqrt{1+\f{\beta^2\hbar^2\omega^2\mu^2}{4}}\right]+\f{\lam^2}{2\mu\omega^2},\nonumber 
\\
\psi_n(p) = N_n e^{-\f{\lam \,{\rm tan}^{-1}(\sqrt{\b}p)}{\hbar \mu\omega^2\sqrt{\b}}}\left(1+\beta p^2\right)^{-\big(\f{\gamma}{2\b}+\f{A}{\sqrt{\b}}\big)} P_n^{\big(\f{A}{\sqrt{\b}}-\f{1}{2},
\f{A}{\sqrt{\b}}-\f{1}{2}\big)}\left(\f{\sqrt{\b}p}{1+\b p^2}\right).\label{wf}
\end{gather}
Thus we find that the spectrum is completely real and for $\lam=0$ it reduces to the known re\-sults~\cite{kempf2,kempf3,dadic,gemba}.

\section{Swanson model}\label{swanson}

We now consider another type of model, namely, the Swanson model with the Hamiltonian given by \cite{swanson}
\begin{gather}
H = \omega a^\dagger a + \lam a^2 + \d{a^\dagger}^2 + \f{\om}{2},\label{swan1}
\end{gather}
where $\lam\neq \d$ are real numbers and $a$, $a^\dagger$ are annihilation and creation operators of the standard harmonic oscillator. Although the above Hamiltonian involves no complex coupling it is non-Hermitian and has real eigenvalues provided $(\om^2-4\lam\d)>0$ \cite{swanson}.

We shall now obtain exact solutions of the Swanson model in the presence of a minimal length. In this case the operators $a$, $a^\dagger$ are defined exactly as in the standard case except that~${\hat x}$ and~${\hat p}$ are given by (\ref{rep1}):
\begin{gather*}
a = \f{1}{\sqrt{2m\hbar \omega}}\left({\hat p}-i\omega{\hat x}\right),\qquad a^\dagger = \f{1}{\sqrt{2m\hbar \omega}}\left({\hat p}+i\omega{\hat x}\right).
\end{gather*}
Now using (\ref{scalar1}) it can be shown that $H\neq H^\dagger$ so that the Hamiltonian (\ref{swan1}) is non-Hermitian.

In order to obtain the spectrum we now write the eigenvalue equation $H\psi(p)=E\psi(p)$ in momentum space as
\begin{gather}
H\psi(p) = \left[-f(p) \f{d^2}{dp^2} + g(p)\f{d}{dp}+h(p)\right]\psi(p) = \eps \psi(p),\label{swan2}
\end{gather}
where $f(p)$, $g(p)$, $h(p)$ and $\eps$ are now given by
\begin{gather*}
f(p)  =  \big(1+\beta p^2\big)^2,\\
g(p)  = {-2\left[\f{2(\delta-\lambda)}{\hbar m\omega(\omega-\lambda-\delta)}+2(\b+\gamma)\right]\big(1+\beta p^2\big)p},\\
h(p) = {\left[ \f{\omega+\lambda+\delta}{\omega-\lambda-\delta} \f{1}{m^2\hbar^2\omega^2}
-\f{2\gamma(\delta-\lambda)}{(\omega-\lambda-\delta)\hbar m\omega}-\g^2\right]p^2}\\
\phantom{h(p) =}{} -\left[\f{\d-\lam+\om}{\hbar m\om(\om-\lam-\d)}+\g\right]\big(1+\b p^2\big),\\
\eps ={\f{1}{{\hbar}m(\om-\lam-\d)}\left(\f{2E}{\om}-1\right)}. 
\end{gather*}
Now performing the transformation (\ref{t}) we obtain from (\ref{swan2})
\begin{gather*}
\left[-\f{d^2}{dq^2}+V(q)\right]\phi(q) = \eps\phi(q),
\end{gather*}
where the potential is given by
\begin{gather}
V(q) = \n \,{\rm  sec}^2(\sqrt{\b}q) + \f{4\lam\d-\om^2}{\hbar^2m^2\om^2\b(\om-\d-\lam)^2},\label{v2}
\\
\nu =  {\f{\om^2-4\lam\d-\hbar m\om^2\b(\om-\d-\lam)}{\hbar^2m^2\om^2\b(\om-\d-\lam)^2}}.\nonumber
\end{gather}
Now proceeding as before the energy eigenvalues and the eigenfunctions are found to be
\begin{gather}
E_n=\frac{\hbar m \omega \beta(\omega-\lambda-\delta)}{2}\left(n^2+n+\frac{1}{2}\right) +\left(n+\frac{1}{2}\right)\sqrt{\left[\omega-\frac{\hbar m \omega \beta(\omega\!-\!\lambda\!-\!\delta)}{2}\right]^2\!-4\lambda \delta},\!\!\label{ener2}
\\
\psi_n(p) = N_n\big(1+\beta p^2\big)^\kappa~P_n^{(s,s)}\left(\f{\sqrt{\b}p}{1+\b p^2}\right),\qquad n=0,1,2,\dots,\nonumber 
\end{gather}
where
\begin{gather*}
s = \frac{\sqrt{1+\frac{4\nu}{\beta}}}{2},\qquad
\kappa = \frac{\lambda-\delta}{2\hbar m \omega}(\omega-\lambda-\delta)-\frac{\gamma}{2\beta}-\frac{1+\sqrt{1+\frac{4\nu}{\beta}}}{2}.
\end{gather*}
From (\ref{ener2}) it follows that the energy is real provided
\begin{gather}
\left[\omega-\frac{\hbar m \omega \beta(\omega-\lambda-\delta)}{2}\right]^2-4\lam\d>0\label{constraint1}
\end{gather}
and for $\b=0$ we recover the standard Swanson model constraint mentioned earlier. From (\ref{constraint1}) it also follows that for given $\omega$, $\lam$, $\d$ (such that $\om-2\sqrt{\lam\d}>0$) there is a critical value $\b_c$ such that for $\b<\b_c$ the energy is real. This value is given by
\begin{gather}
\b_c = \f{2(\om-2\sqrt{\lam\d})}{m{\hbar}\om(\om-\lam-\d)}.\label{constraint}
\end{gather}
Thus in this case apart from the standard Swanson model constraint, there is an additional constraint (\ref{constraint}) involving the minimal length parameter.

\section[$\eta$ pseudo-Hermiticity]{$\boldsymbol{\eta}$ pseudo-Hermiticity}\label{pseudo}

We recall that a Hamiltonian $H$ is called $\eta$ pseudo-Hermitian if it satisfies the condition \cite{mostafa}
\begin{gather*}
\eta H\eta^{-1} = H^\dagger,
\end{gather*}
where $\eta$ is a Hermitian operator. It may be noted that for $\eta$ pseudo-Hermitian systems the usual scalar product (\ref{scalar1}) can not be used since it may lead to a norm with f\/luctuating sign. The scalar product for such systems is def\/ined as
\begin{gather}
\langle \phi(p)|\psi(p)\rangle_{\eta} = \langle \phi(p)|\eta\psi(p)\rangle . \label{scalar2}
\end{gather}
Thus in the present case scalar product reads
\begin{gather*}
\langle\phi(p)|\psi(p)\rangle_{\eta}=\int \f{\eta\phi^*(p)\psi(p)}{(1+\beta p^2)^{1-\f{\gamma}{\beta}}}\,dp.
\end{gather*}
Also $\eta$ pseudo-Hermitian systems are characterized by the fact that their spectrum is either completely real or the eigenvalues occur in complex conjugate pairs \cite{mostafa}. Since in both the models considered here the eigenvalues are real it is natural to look for $\eta$ pseudo-Hermiticity of the Hamiltonians (\ref{sch1}) and (\ref{swan1}).

Next we take the metric as
\begin{gather}
\eta = \big(1+\b p^2\big)^{-\f{\g}{\b}}\exp\left[-\int(\chi+\chi^*)\,dp\right].\label{metric}
\end{gather}
Then using (\ref{fgh}) and (\ref{rho}) the metric for
the displaced oscillator is found to be
\begin{gather}
\eta_{ho} = \exp\left[\f{2\lam}{\hbar\mu\sqrt{\beta}\omega^2}{\,{\rm tan}^{-1}\big(\sqrt{\beta}p\big)}\right].\label{eta1}
\end{gather}
Now it can be shown that (\ref{eta1}) satisf\/ies
\begin{gather}
\eta_{ho} H\eta_{ho}^{-1} = H^\dagger,\label{prop1}
\end{gather}
so that $H$ is $\eta$ pseudo-Hermitian. It can also be verif\/ied that the wave functions (\ref{wf}) are orthonormal with respect to the scalar product (\ref{scalar2}):
\begin{gather}
\langle\psi_m(p)|\psi_n(p)\rangle_{\eta} = \delta_{mn}.\label{prop2}
\end{gather}
Similarly using (\ref{metric}) the metric for the Swanson model can be found to be
\begin{gather*}
\eta_s = \left(1+\b p^2\right)^{\f{(\d-\lam)}{\hbar m\om\b(\om-\lam-\d)}}.
\end{gather*}
It can be verif\/ied that the Swanson Hamiltonian (\ref{swan1}) satisf\/ies the relations (\ref{prop1}) and (\ref{prop2}). Thus the Swanson model is also $\eta$ pseudo-Hermitian.

\section{Discussion}\label{con}
In this paper we have obtained exact solutions of a couple of non-Hermitian models in a space admitting a minimal length. In the case of the displaced oscillator the spectrum is real irrespective of the coupling strength and for the Swanson model the spectrum is real subject to certain constraints on the parameters.
In this context we note that non-Hermiticity can also be introduced in a model by considering non-Hermitian coordinates, i.e.~${\hat x}^\dagger\neq {\hat x}$. This may be achieved by replacing ${\hat x}\ra {\hat X}={\hat x} + i\eps$ so that ${\hat X}\neq {\hat X}^\dagger$. However this case reduces to the model considered in Section~\ref{dho} once the parameters $\eps$ and $\lam$ are suitably related. A second possibility is to consider replacing $\g$ by $i\g$ in (\ref{rep1}). With such a replacement the harmonic oscillator Hamiltonian becomes non-Hermitian although the spectrum will still remain real. We would now like to mention about the symmetry of the problems considered here. Since the transformation (\ref{t}) of the variable $p$ to the variable $q$ is invertible, it is expected that the symmetry of the original problem is the same as that of the corresponding Schr\"odinger one~\cite{kamran}. Since the underlying symmetry of the potentials~(\ref{pot1}) and~(\ref{v2}) is a nonlinear algebraic one~\cite{quesne} we expect that the original problems to have the same symmetry. We feel it would be interesting to investigate the symmetry structure of these types of models.

Finally in view of the fact that the representation of the position operator in higher dimension is non trivial we feel it would be interesting to examine non-Hermitian interactions in higher dimensions and also to examine solvability of Schr\"odinger equation with other types of non-Hermitian interactions.

\subsection*{Acknowledgments}

The authors would like to thank the referees for suggesting improvements.

\pdfbookmark[1]{References}{ref}
\LastPageEnding

\end{document}